\documentclass[a4paper,11pt]{article}
\usepackage{amsmath,amssymb,amsthm,bbm,bm}  
\usepackage{mathrsfs}
\usepackage{authblk}  
\usepackage[british]{babel}
\usepackage{balance}
\usepackage[utf8]{inputenc}
\usepackage[backend=biber,backref=true,sorting=none,doi=false,isbn=false,giveninits=true,style=nature,url=false]{biblatex} 
\usepackage{booktabs} 
\usepackage{caption}
\usepackage{color}
\usepackage{comment}
\usepackage{dsfont}
\usepackage{enumitem}
\usepackage[T1]{fontenc}
\usepackage[top=2.5cm, bottom=2.5cm, left=2.5cm, right=2.5cm]{geometry}
\usepackage{graphicx}
\usepackage{graphbox}       
\PassOptionsToPackage{hyphens}{url}  
\usepackage[colorlinks = true,
            linkcolor = OliveGreen,
            urlcolor  = blue,
            citecolor = MidnightBlue,
            anchorcolor = blue]{hyperref}
\usepackage{cleveref}
\usepackage{csquotes}  
\usepackage[mono=false]{libertine}  
\usepackage{mathtools}
\usepackage{microtype}      
\usepackage{multirow}    
\usepackage{nicefrac}
\usepackage{physics}
\usepackage{subfigure}
\usepackage{wrapfig}
\usepackage{float}
\usepackage{mathtools}

\usepackage[usenames,dvipsnames]{xcolor}

\newcommand{\citet}[1]{\textcite{#1}}
\newcommand{\citep}[1]{\parencite{#1}}

\newcommand{\identity}{\text{Id}}
\newcommand{\stdGauss}{{\mathcal{N}\left(0,\identity\right)}}

\global\long\def\identity{\mathrm{Id}}%
\global\long\def\stdGauss{\mathcal{N}\left(0,\identity\right)}%
\global\long\def\abt{\bar{\alpha_{t}}}%

\newtheorem{theorem}{Theorem}
\newtheorem{lemma}[theorem]{Lemma}

\bibliography{references}

\title{Local Coverage Governs Memorization in Diffusion
Models}

\author{Claudia Merger}
\author{Sebastian Goldt\thanks{\{cmerger, sgoldt\}@sissa.it}}
\affil{International School of Advanced Studies (SISSA), Trieste, Italy}

\date{\today}

\begin{document}

\maketitle

\begin{abstract}
Memorization in diffusion models is often treated as a global property of the
model or dataset. In practice, however, a single diffusion model can
simultaneously generate both memorized and novel samples. Which training samples
are most likely to be memorized?
In this work, we show that memorization is governed by \emph{local data
coverage}. Leveraging the connection between diffusion models and kernel density
estimation (KDE), we derive a theoretical criterion that predicts whether a
point is memorized based on the density of training data in its
neighborhood and the size of the training dataset. In the high-dimensional
limit, this leads to a sharp, local transition: regions of low coverage are
dominated by isolated training samples, which are memorized, while dense regions
support interpolation and generalization.
We validate these predictions empirically, showing that memorization increases with local sparsity and that diffusion models exhibit a coexistence of memorized and novel samples within the same model. Extending this framework to multi-class settings, we further show that classes with higher intra-class diversity (and thus lower local coverage) are more strongly memorized. 
Our results provide a local view of memorization in diffusion models, explaining when and where memorization occurs in terms of data geometry.
\end{abstract}

\section{Introduction}

\paragraph{How does memorization arise \emph{locally} in diffusion models?}
Diffusion models are powerful generative methods capable of learning complex
high-dimensional data distributions from finite datasets. Yet when trained with
limited data, they may reproduce training examples rather than generate novel
samples, a phenomenon commonly referred to as
\emph{memorization}~\citep{somepalli_diffusion_2022, carlini_extracting_2023,
kadkhodaie2024generalization}. Understanding the mechanisms that drive
memorization is important because memorized samples can lead to unintended data
leakage, privacy violations, and the release of copyrighted or sensitive
training data.

\begin{figure*}[!t]
    \centering
    \includegraphics[width=\linewidth]{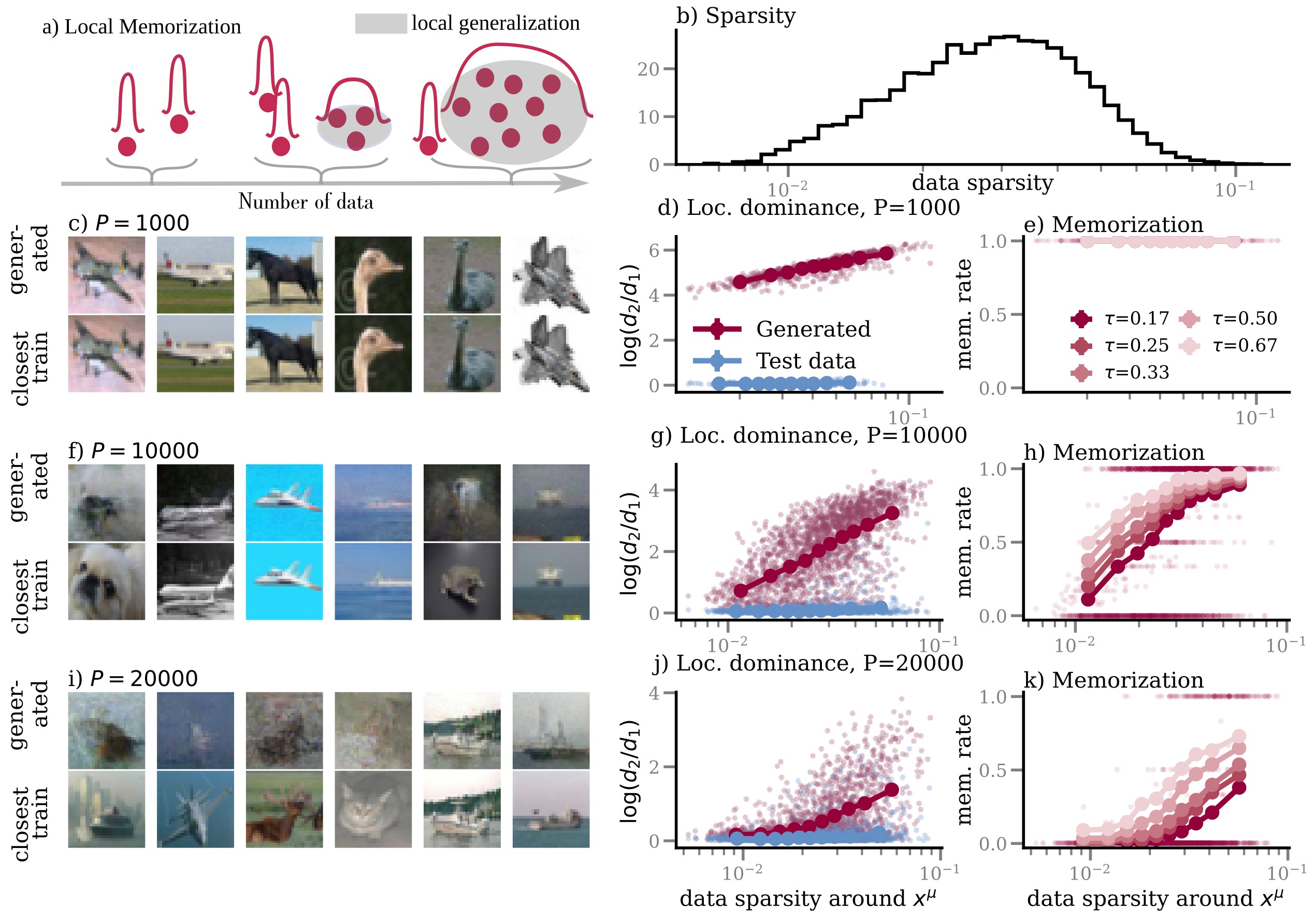}
    \caption{\label{fig:fig1} \textbf{Local data coverage governs memorization.}
    a) Sketch of local memorization in kernel density approximation.
    b) Histogram of data sparsity, measured by the average squared distance of a
    training point its 10 nearest neighbors, for $20{,}000$ CIFAR10 samples.
    c) Generated samples + closest training image (measured by $\ell_2$ distance) for a diffusion model trained on $P=1000$ training examples. 
    d) For each generated sample $\hat{x}$,
    we compute the local dominance as the log-ratio $\log d_2 / d_1$ between
    the distances $d_1$ and $d_2$ to its nearest and second-nearest
    training samples. For each training sample $x^\mu$, we show distances to
    generated images (red) and test data (blue), providing a
    baseline without memorization. Scatter shows raw values, markers indicate binned averages.
    e) Memorization rate per training sample $x^\mu$, defined as the fraction of generated samples satisfying $d_1/d_2 < \tau$. Scatter shows raw values for $\tau=0.1$, markers indicate binned averages across thresholds.
    f)--k) Same measurements for models trained on larger datasets. As the number of training samples increases, both dominance and memorization decrease, and their dependence on sparsity weakens.
    Overall, sparse regions exhibit strong single-sample dominance and higher memorization, while dense regions promote interpolation across multiple training points.}
\end{figure*}

Theoretical work has begun to analyze memorization in linear
models~\citep{maillard2026memorisation, wang2026random}, in random feature
models~\citep{george2026denoising, bonnaire2026diffusion}, and by exploiting the
close connection between diffusion models and kernel density estimation (KDE)
\citep{pidstrigach_score-based_2022, pham_memorization_2024, ambrogioni_search_2024, biroli_kernel_2024,
achilli_capacity_2025, li_good_2024, lucibello_exponential_2024}. In the latter picture, each
training sample contributes a local kernel, and generalisation arises from the
superposition of many such kernels, see \cref{fig:fig1}a for a sketch. When
kernels overlap strongly, the model assigns finite probability to the space
between examples; when overlap is weak, generation can collapse onto individual
training points. These analyses predict a \emph{global} phase transition: the whole generative model goes
from memorizing to generating new samples at once.

Recent work has begun to paint a more fine-grained picture of memorization. In
practice, it was observed that models often reproduce only a subset of their
training data rather than the entire dataset~\citep{fang_closer_2025,
carlini_extracting_2023}; the effect depends strongly on the training setup
(see further related work below). Thus, at finite sample size, a diffusion model
often generates a mixture of copied and novel samples see \cref{fig:fig1} c) f)
and i). On the theoretical side, \citet{achilli_losing_2026} showed that
memorization may emerge progressively through the loss of manifold dimensions,
leading to a form of \emph{geometric memorization} in which some directions of
variability are lost before exact copying occurs.
\citet{garnier-brun_biased_2026} showed a coexistence of memorization and
generalization in early-stopped diffusion models.

Taken together, these findings suggest that memorization is richer than a single
global transition and raise a natural question:
\begin{center}
\textbf{Which regions of data space, classes, or individual samples\\ are most likely to be memorized?}
\end{center}
In this work, we give a concrete criterion for memorization by introducing a
\emph{local} theory of memorization in diffusion models. Our central hypothesis
is simple: memorization is governed not only by the total number of training
samples, but by their \emph{local coverage}. Regions of data space containing
many nearby examples support interpolation and generalization, whereas isolated
regions are prone to sample retrieval. In \cref{fig:fig1} we show an example of
this behavior for diffusion models with U-net architecture
\citep{ronneberger_u-net_2015} trained on subsets of the CIFAR10 image dataset \citep{krizhevsky_learning_2009} of increasing size $P$. When trained on very few data points, the models first produce only memorized data. As we increase the number of training data, the models begin to generate a mixture of memorized and novel examples (panels c,f,i). We then quantify to which degree samples generated from diffusion models replicate the closest training data (panels d,g,j) as a function of local sparsity. We find isolated training points in low density (high sparsity) regions are preferentially memorized (panels e,h,k), whereas samples generated in regions of higher local density (lower local sparsity in \cref{fig:fig1}) are less likely to be memorized. See \cref{sec:experiments} for full experimental details.

\paragraph{A local theory of memorization} To formalize this intuition, we
develop a local theory of memorization using the KDE framework of
\citet{lucibello_exponential_2024, biroli_kernel_2024,
achilli_memorization_2025}. Here, we sketch the main idea. We quantify the local
coverage around a point~$x \in \mathbb{R}^N$ by the probability mass contained
in a ball of radius $h$ around $x$,
\begin{equation}
p_{\mathrm{in}}(x)=\int_{B_h(x)} d\rho(x'),
\end{equation}
where $B_h(x)$ denotes a $N$-dimensional hypersphere centered at $x$ and $\rho$
the density of the data from which the training points are drawn. This quantity
measures how densely the data distribution $\rho$ populates the neighborhood of
$x$. In the high-dimensional regime in which the number of samples scales
as~\citep{lucibello_exponential_2024, biroli_kernel_2024}
\begin{equation}
\textstyle
    P=e^{\alpha N} \,,
\end{equation} we define 
\begin{equation}
\textstyle
    \nu_{\text{in}}(x) = \lim_{N\rightarrow \infty} \frac{1}{N}\ln p_{\text{in}}(x)\,. \label{eq:nu_definition}
\end{equation}
Our main result shows that memorization at a point $x$ is controlled by the pair $(\nu_{\mathrm{in}}(x),\alpha)$ with a sharp transition: 
\begin{equation}
    \begin{cases}
        x \in A_{\text{mem}} & \text{if } \nu_{\text{in}}(x)<-\alpha \\
        x \in A_{\text{gen}} & \text{otherwise}
    \end{cases}
    \label{eq:RegionSeparation}
\end{equation}
where we refer to $A_{\text{gen}}$ as the region where the model correctly interpolates, and $A_{\text{mem}}$ is a region where the learned density is characterized by isolated peaks centered on training examples. 
This result shows that depending on local coverage, the same diffusion model may
exhibit retrieval-like behavior in some regions of space and generative
interpolation in others. Points in low-coverage regions, where
$\nu_{\text{in}}(x)$ is small, are memorized, while points in high-coverage regions are generalized.

\paragraph{Contributions} In the following, we recall the relation between diffusion
models and kernel density estimators in \cref{sec:diffusion_models}. Our main
contributions are then as follows
\begin{itemize}
    \item We develop \textbf{a local theory of memorization} and prove that the
    transition from memorization to generalization in a point is governed by the
    local kernel statistic $\nu_\mathrm{in}$, see \cref{thm:W_regimes} and
    \cref{sec:theory}.
    \item We show that the theory explains several empirical properties of
    diffusion models trained on CIFAR10 and CelebA \citep{liu_deep_2015} in
    \cref{sec:experiments}:
    \begin{itemize}
    \item \textbf{Coexistence of memorization and generalization:} the same
    model can simultaneously generate copied and novel samples; as $P$
    increases, memorization vanishes first in dense regions and persists longest
    in sparse regions, see \cref{fig:fig1} and \cref{sec:local-sparsity}. 
    \item \textbf{Class-dependent memorization:} classes with larger intra-class
    diversity (and therefore lower local coverage) are memorized more strongly,
    see \cref{fig:classwise_cifar} and \cref{sec:class-wise-memorisation}.
    \end{itemize}
\end{itemize}

More broadly, our results suggest that memorization in diffusion models is not a
global binary property, but a local finite-sample phenomenon governed by the
geometry and coverage of the training distribution. This mechanistic insight
into memorization opens up the possibility of designing dedicated strategies to  both 
suppress unwanted memorization and to
retain specific training samples.

\clearpage

\subsection*{Further related Work}

\paragraph{Impact of training setup on memorization} A growing body of empirical
and theoretical work has established that memorization depends strongly on
properties of the training setup. In particular, it is exacerbated by data
duplication~\citep{carlini_extracting_2023}, prompt conditioning and
classifier-free guidance~\citep{somepalli_understanding_2023,
wen_detecting_2024, kim_how_2025, gu_memorization_2025}, model
capacity~\citep{yoon_diffusion_2023, george_denoising_2025,
zhang_generalization_2025}, dataset
structure~\citep{somepalli_understanding_2023, kim_how_2025,
gu_memorization_2025, yoon_diffusion_2023}, and training
duration~\citep{bonnaire2026diffusion, favero_bigger_2025,
garnier-brun_biased_2026}, while decreasing with learning
rate~\citep{wu_taking_2025} and dataset
size~\citep{somepalli_understanding_2023, kadkhodaie2024generalization,
maillard2026memorisation}.

\paragraph{Local density estimators and memorization in practice} Strategies to
measure and mitigate memorization include determining the local intrinsic
dimension of generated points \citep{ross_geometric_2025}, and estimating how
``peaked'' the estimated density is around a given point
\citep{jeon_understanding_2025} as well as identifying memorized
prompts~\citep{wen_detecting_2024, kim_how_2025}. These strategies share a
common intuition: isolated data points produce a highly peaked local density,
and produce samples of low intrinsic dimension. This perspective is naturally
consistent with a kernel density estimation (KDE) view of diffusion models.
Isolated training points induce highly concentrated local density estimates and
generate low-dimensional samples, while dense regions support smoother
interpolation. Similarly, conditioning can be interpreted as restricting the
effective sample set contributing to the density estimate, thereby increasing
local sparsity and promoting memorization. In this sense, existing empirical
methods already implicitly rely on a local density perspective on memorization.

\paragraph{Memorisation in kernel density estimation} On the theoretical side, several works have established a connection between
diffusion models and kernel density estimation
\citep{pidstrigach_score-based_2022, pham_memorization_2024, ambrogioni_search_2024, biroli_kernel_2024,
achilli_capacity_2025, li_good_2024, lucibello_exponential_2024}.
Within this framework, one can determine a global ``collapse phase'' in
diffusion models as a function of effective noise in the backward/sampling
process \citep{biroli_dynamical_2024}, where the model begins to reproduce
training data. More generally, the sample complexity required for smooth
interpolation with kernel density estimators is known to scale exponentially
with the ambient or manifold dimension\citep{lucibello_exponential_2024,
achilli_memorization_2025} of the data. Recent work has further shown that
memorization can emerge gradually through a loss of manifold dimensions
\citet{achilli_losing_2026} and that this process may be spatially
inhomogeneous. However, existing theoretical analyses remain largely global:
they characterize when a model as a whole memorizes or generalizes, but do not
predict which specific regions or samples are more likely to be memorized. In
contrast, we develop a local theory of memorization that explicitly links
memorization to data coverage at the level of individual points. This theory provides a
principled explanation for the heterogeneous memorization patterns observed in
practice.

\section{Background: from diffusion models to kernel density estimation}%
\label{sec:diffusion_models}

We briefly recall a key observation: diffusion models trained on finite data
behave like kernel density estimators (KDE). 
Diffusion models define a forward noising process that gradually corrupts a data point $x_0$ from $\rho$ through the addition of noise 
\begin{equation}
    x_{t} \left(\epsilon_t \right)= 
    \sqrt{\bar{\alpha}_{t}}x_{0}
    +\sqrt{1-\bar{\alpha}_{t}}\epsilon_{t}
    ,\qquad\epsilon_{t}\sim\mathcal{N}\left(0,\identity\right)\,
    ,\label{eq:x_t_from_x_0}
\end{equation}
where $\bar{\alpha}_{t}\in(0,1)$ is a decreasing function in $t$. As $t$ increases, the original signal $x_0$ is gradually suppressed compared to the isotropic Gaussian noise, until one obtains $x_T$ whose distribution is close to $\stdGauss$. A neural network $\epsilon_{\theta}$ with parameters $\theta$ is trained to predict the noise $\epsilon_t$, which corresponds to learning the score $\nabla\ln \rho_t(x_t)$ , where $\rho_t$ is the distribution of $x_t$. This is achieved via the denoising score matching objective 
\begin{equation}
\textstyle
    L = \frac{1}{N\,T P} \sum_t \sum_{x^{\mu} \in \mathcal{D}}
    \mathbb{E}_{\epsilon_t}\left\Vert 
     \epsilon_t 
     -\epsilon_{\theta}
      \left(
       x^{\mu}_{t} \left(\epsilon_t \right),t
      \right)
    \right\Vert^{2}\,,  
    \label{eq:loss}
\end{equation}
where $\mathcal{D}$ is a training set of size $P$ points $\{ x^{\mu}\}_{\mu=1}^P$ drawn i.i.d. from $\rho$. 

To make the connection to KDE explicit, consider an idealized setting in which the model has infinite capacity and is trained to optimality. In this case, for each $t$, the learned function $\epsilon_{\theta}(\cdot,t)$ can be treated as an arbitrary function $\epsilon(\cdot,t)$ and the loss can be minimized functionally. Setting this functional derivative to zero yields:
\begin{equation}
\textstyle
    \epsilon(x,t) \propto \nabla_x \ln \sum_{\mu} \mathcal{N} \left(x; \abt x^{\mu}, (1-\abt) \identity  \right) \,.
    \label{eq:KDEscore}
\end{equation}
This identity reveals that, when trained to convergence on a finite number of samples, diffusion models learn a Gaussian mixture centered at the training data points.  In other words, diffusion models trained on a finite dataset recover the score of the empirical distribution convolved with Gaussian noise, rather than the true underlying density $\rho$. In the low-noise limit ($\abt \rightarrow 1$), this mixture becomes sharply peaked around individual training samples, and the model effectively reproduces them.
At finite noise, there is a direct correspondence between \cref{eq:KDEscore} and
KDE. KDE attempts to approximate a density $\rho$ using a mixture of kernel
functions $K_N$, indexed by the input dimension $N$, centered at the data points:
\begin{equation}
\textstyle
\label{eq:Z}
    Z_N(x) = \frac{1}{P} \sum_{\mu} K_N\left( \frac{\left| x-x^{\mu}\right|}{h} \right) \,.
\end{equation}
The correspondence with KDE becomes explicit by identifying the kernel as Gaussian with bandwidth $h=\sqrt{1-\abt}$ up to a rescaling of the variables $x, x^{\mu}$ by $\sqrt{\abt}$. In the ideal case, one has sufficient number of samples $P$ such that $Z$ becomes a smooth approximation of $\rho$. This behavior with $P$ is one hypothesis to explain memorization and generalization in diffusion models.

This perspective suggests that generalization is not uniform across the data space. Even when the total number of samples is large, the quality of the KDE approximation depends on local sample density. Regions with high data density are well approximated, while sparse regions remain dominated by individual kernels. In the following, we formalize this intuition by deriving conditions under which local regions exhibit memorization or generalization.

\section{Theory: Kernel density estimation with hard spheres}
\label{sec:theory}

\subsection{General theory and main result}

To study kernel density estimation (KDE), we characterize the behavior of the log-density
\begin{equation}
    z_N(x_N) = \frac{1}{N}\ln Z_N(x_N)\,.
    \label{eq:def_z}
\end{equation}
where $Z_N(x)$ is the mixture of kernel functions $K$ centered at the data points defined in \cref{eq:Z}. The difficulty in establishing the behavior of $z_N$ is that it is a random quantity that depends on the draw of the $P$ training points from $\rho$. Our goal is to compute its distribution over such draws to derive general properties of high dimensional KDE. 

Since the ambient dimension changes with $N$, the symbol $x_N$ should not be understood as a fixed point in a fixed space. We instead consider a sequence of vectors $\mathbf x=(x_N)_{N\geq 1}$ and distributions $\rho_N$ with $x_N\in\mathbb R^N$, and assume that the corresponding local statistics have a limit. Equivalently, the result is indexed not by the coordinate vector $x_N$ itself, but by the limiting local environment around it, namely the distribution of
\[
    \epsilon_{\mathbf x,N}(\tilde x_N)
    =
    \frac1N \ln K_N\!\left(\frac{x_N-\tilde x_N}{h}\right),
    \qquad \tilde x_N\sim\rho_N,
\]
Thus, the behavior of $z_N(x_N)$ is controlled by the distribution of the scalar statistic $\epsilon_{
\mathbf{x},N}$ rather than by the full high-dimensional variable $\tilde x_N \sim \rho_N$.

We use an exponential transform, inspired by \citet{dotsenko_replica_2011}, to compute the cumulative distribution function of $z_N(x_N)$. Define
\begin{equation}
    \overline{W_N(y,x_N)}
    =
    \overline{
    \exp\left(
    -\exp\left[-N\left(y-z_N(x_N)\right)\right]
    \right)}
    \,,
    \label{eq:W_N_definition}
\end{equation}
where the overline denotes the average over draws of the $P$ training samples. As $N\to\infty$, the inner function converges pointwise to $e^{-e^{-N(\cdot)}}\rightarrow\Theta(\cdot)$, with convention $\Theta(0)=e^{-1}$. Hence if the limiting CDF $F_{\mathbf{x}}(y) = \lim_{N\rightarrow \infty} \Pr(z_N(x_N)\leq y)$ is continuous in $y$, then $\overline{W_N(y,x_N)}$ converges to $F$ as $N\rightarrow \infty$; see \cref{app:convergence}.

Using independence of the training samples, the average factorizes and gives
\begin{equation}
    \overline{W_N(y,x_N)}
    =
    \exp\left\{
    e^{\alpha N}
    \ln\left[1-I_N(y,x_N)\right]
    \right\},
    \label{eq:W_N_averaged}
\end{equation}
with
\begin{equation}
    I_N(y,x_N)
    =
    \mathbb E_{\tilde x\sim \rho_N}
    \left[
    1-
    e^{
    -e^{
    N\left[\epsilon_{x,N}(\tilde x)-\alpha-y\right]
    }
    }
    \right].
    \label{eq:I_N_kernel_statistic}
\end{equation}
This yields the following theorem, which is our main theoretical result:

\begin{theorem}[Cumulative distribution function of KDE]
\label{thm:W_regimes}
Assume that the cumulative distribution function converges $F_{\mathbf{x},N}(y) = \Pr(z_N(x_N)\leq y) \to F_{\mathbf{x}}(y)$ for all $y$ as $N\rightarrow\infty$. Furthermore, let us assume that the limit
\begin{equation}
    i_{\mathbf{x}}(y)
    :=
    \lim_{N\to\infty} e^{\alpha N} I_N(y,x_N)
    \label{eq:i_x}
\end{equation}
exists with $i_{\mathbf{x}}(y) \in [0,\infty]$. Then if $F_{\mathbf{x}}(y)$ is continuous in $y$, we find
\begin{equation}
    F_{\mathbf{x}}(y)
    =
    \begin{cases}
        0 & \text{if  } i_{\mathbf{x}}(y)=\infty,\\
        \exp[-i_{\mathbf{x}}(y)] & \text{if }\, 0<i_{\mathbf{x}}(y)<\infty,\\
        1 & \text{if  } i_{\mathbf{x}}(y)=0.
    \end{cases}
    \label{eq:W_scaling}
\end{equation}
\end{theorem}
We detail the proof in \cref{app: Derivations}. 

Theorem \ref{thm:W_regimes} shows that the limiting CDF of $z_N(x_N)$ is
determined by the exponential scaling of the probability that the kernel
statistic $\epsilon_{\mathbf{x},N}$ is large enough to contribute at level $y$.
In typical high-dimensional settings, $z_N(x_N)$ concentrates, so the
intermediate regime in \cref{eq:W_scaling} collapses to the location of a sharp
jump. Fluctuations of $\epsilon_{\mathbf{x}}$ would instead broaden this
transition.

\paragraph{Relation to Gumbel's law.} The form of \cref{eq:W_N_averaged} is somewhat reminiscent of Gumbel's law which computes the cumulative distribution function of the maximum of $P$ i.i.d. random variables, in our case, this amounts to replacing $z(x)$ by $\tilde{z}(x)=\ln \max_{\mu} K\left(\frac{x-x^{\mu}}{h}\right)/P$. This structural similarity is no accident, indeed one can show that whenever $z(x)\approx \tilde{z}(x) $, \cref{eq:W_N_averaged} directly corresponds to Gumbel's law. In the generalization regime however, exponentially many terms contribute to $z$, and the analogy breaks down. We discuss this relation in more detail in \cref{app:gumbel_relation}. 

\subsection{Case study: hard-sphere kernels}

We now specialize to the hard-sphere kernel
\begin{equation}
    K_N(u)=\frac{\Theta(R^2-\|u\|^2)}{V_R},
    \label{eq:hard_sphere_kernel}
\end{equation}
where $V_R$ is the volume of the $N$-dimensional ball of radius $R$.  In high
dimensions, the hard-sphere kernel captures the same radial concentration
mechanism as a Gaussian kernel. At the level of the exponential transform, the
relevant quantity is the distribution of $\epsilon_{x,N}(\tilde x_N)$. For a
Gaussian kernel, this statistic is, up to additive constants, proportional to $
-\frac{1}{N}\|x_N-\tilde x_N\|^2 .$ Thus the leading exponential behavior is
governed by level sets of the squared distance. In high dimensions, the Gaussian
measure concentrates on a thin shell at its typical radius. Similarly, for the
hard-sphere kernel, almost all of the volume of the ball concentrates near its
boundary. Therefore, while Gaussian and hard-sphere kernels are quantitatively
different, the hard-sphere kernel provides a tractable model that exposes the
same local-coverage transition mechanism. For simplicity, we  fix $R = 2\pi e$, for which the volume $V_R$ remains $\mathcal{O}(1)$ as $N \to \infty$. The effective scale of the kernel relative to the data is controlled by the bandwidth $h$, which can be interpreted as a rescaling of the data points by a factor $1/h$. Unless otherwise specified, we set $h = 1$.

In this case, we find that the quantity controlling the limiting behavior is the local coverage. Let us assume that the limit of the local coverage rate
\begin{equation}
    \nu_{\mathrm{in},\mathbf x}
    =
    \lim_{N\to\infty}\frac1N\ln p_{\mathrm{in},N}(x_N).
    \label{eq:nu_in_def}
\end{equation}
exists. Then we obtain the following result:
\begin{lemma}[Local hard-sphere KDE transition]
\label{lemma:local_kde_transition}
Let $P=e^{\alpha N}$ with fixed $\alpha>0$, and let $K_N$ be the hard-sphere kernel in \cref{eq:hard_sphere_kernel}. Under the assumptions of \cref{thm:W_regimes} and additionally assuming that that $\nu_{\mathrm{in},\mathbf x}$ exists, when $\nu_{\mathrm{in},\mathbf x}\neq -\alpha$ we find
\begin{equation}
    z_N(x_N)
    \xrightarrow[N\to\infty]{\mathbb P}
    \begin{cases}
        \nu_{\mathrm{in},\mathbf x} & \text{if } \nu_{\mathrm{in},\mathbf x}>-\alpha,\\
        -\infty & \text{if } \nu_{\mathrm{in},\mathbf x}<-\alpha.
    \end{cases}
    \label{eq:z_scaling_theorem}
\end{equation}
\end{lemma}
The proof follows from a direct application of \cref{thm:W_regimes}, see \cref{app: Derivations}.

Lemma \ref{lemma:local_kde_transition} establishes a local transition controlled by the comparison between local coverage $\nu_{\mathrm{in}}(x)$ and sample complexity $\alpha$. Regions with $\nu_{\mathrm{in}}(x)>-\alpha$ contain exponentially many nearby samples and therefore support smooth KDE estimates. Regions with $\nu_{\mathrm{in}}(x)<-\alpha$ contain no nearby samples with high probability; in the diffusion-model interpretation, probability mass assigned to such regions concentrates on isolated training examples, leading to memorization. The divergence $z(x) \to -\infty$ for $\nu_{\text{in}}(x) < -\alpha$ reflects the fact that, in these regions, the kernel density estimate vanishes with high probability. Indeed,
\begin{align*}
\textstyle
    \Pr\left(\left\{ Z(x)=0\right\} \right)	&=\left(1-p_{\text{in}}(x)\right)^{P}  \\
	&=e^{P\ln(1-e^{N\nu_{\text{in},N}(x)})} \\
	&\rightarrow\Theta\left(-\left[\alpha+\nu_{\text{in}, \mathbf{x}}\right]\right)
\end{align*}
On the other hand, when $\alpha$ is large enough, then $Z(x)$ approaches $p_{\text{in}}(x)$, hence the kernel density estimate converges to a local average of $\rho$ over a sphere.

In the context of diffusion models, such regions are dominated by isolated training points. If probability mass is assigned to these locations, it must concentrate on individual samples, leading to memorization. This establishes the coexistence of memorized and generalized regions, as described in Eq.~\eqref{eq:RegionSeparation}.

\section{Experimental results}%
\label{sec:experiments}

\subsection{Local sparsity governs memorization of samples}%
\label{sec:local-sparsity}

We now empirically investigate how local data density influences memorization
behavior in diffusion models. Our goal is to test the central prediction of the
KDE-based theory that sparse regions are more prone to memorization than dense
ones. This implies a coexistence of memorization and generalization: copied and
novel samples can arise simultaneously from the same model. Furthermore, it
predicts a gradual, rather than abrupt disappearance of memorization: as $P$
increases, memorization vanishes first in dense regions and persists longest in
sparse regions. Consequently, regions of low data density lead to more
localized, sample-specific behavior, while dense regions promote interpolation
across multiple training points. While our theory predicts a sharp transition as
a function of local coverage, here we test its qualitative consequences using
empirical proxies for density.

\subsubsection{Experimental methods}

We train diffusion models with a U-net architecture on randomly chosen subsets of CIFAR-10 and CelebA. We then generate samples from these models and assess their closeness to training examples using different measures outlined below. Details on the experimental procedure are given in \cref{app: Details}. We first define how we measure sparsity and memorization and then describe the outcome of the experiments shown in \cref{fig:fig1} and \cref{fig:celeba_local_memo}.

\paragraph{Local sparsity.} For each training sample $x^{\mu}$, we quantify local sparsity using nearest-neighbor distances, which serve as a proxy for inverse local coverage. Concretely, we define
\begin{equation}
\textstyle
    s(x^{\mu}) = \frac{1}{N k} \sum_{i=1}^{k} \|x^{\mu} - x^{\mu_i}\|^2, \label{eq:sparsity_def}
\end{equation}
where $\{x^{\mu_i}\}$ are the $k$ nearest neighbors of $x^{\mu}$. Larger values of $s(x^{\mu})$ correspond to sparser regions of the data distribution. We also validate that the results remain consistent for different choices of $k$ in $\{ 5, 10, 20, 50 \}$. The results reported in the figures correspond to $k=10$. 

\paragraph{Local dominance.} Given a generated sample $\hat{x}$, we compute its distances $d_1(\hat{x})$ and $d_2(\hat{x})$ to the nearest and second-nearest training samples, and define the \emph{dominance}
\begin{equation}
\textstyle
    \Delta(\hat{x}) = \log \frac{d_2(\hat{x})}{d_1(\hat{x})}.
\end{equation}
Large values of $\Delta$ indicate that a single training point dominates the local score estimation around the generated sample, while $\Delta \approx 0$ corresponds to interpolation between multiple neighbors. We assign each generated sample to its nearest training point and compute, for each $x^{\mu}$, the average dominance over all generated samples assigned to it. To disentangle model-specific effects from dataset geometry, we construct a baseline by replacing generated samples with held-out test data, processed in the same way. 

\paragraph{Memorization rate.} We further report a binary memorization metric used in several previous studies \citep{wu_taking_2025, yoon_diffusion_2023,bonnaire2026diffusion} based on the condition $d_1/d_2 < \tau$. 

\subsubsection{Results}

The middle column of \cref{fig:fig1} shows the average dominance as a function of local sparsity for different training set sizes. We observe a clear \textbf{monotonic increase of dominance with sparsity} for generated samples: sparse regions exhibit strong single-sample dominance, while dense regions show low dominance and interpolation. In contrast, the test-data baseline remains nearly flat, indicating that this effect is not explained by nearest-neighbor geometry alone. As the number of training samples increases, the overall level of dominance decreases and the dependence on sparsity weakens, consistent with the expectation that higher sample density reduces isolated regions. The corresponding results for the memorization rate are shown in the right column of \cref{fig:fig1}. While we do not directly estimate the threshold $\nu_{\text{in}}(x) = -\alpha$, the observed dependence on local sparsity is consistent with a transition controlled by local coverage. For the canonical value $\tau=0.33$ commonly used in the literature, we see a clear\textbf{ coexistence of memorization and generalization} that depends on data sparsity. Moreover, while the precise behavior depends on $\tau$, the same qualitative trend is observed: \textbf{memorization increases with sparsity}.  In the appendix \cref{app:additional experiments} we show the outcome of the same experiment on downsampled CelebA, an image dataset of celebrity portraits. The results of these experiments are also consistent with the hypothesis. These results support a local version of the KDE picture: diffusion models exhibit a continuous transition from interpolation in dense regions to single-sample dominance in sparse regions.

\subsection{Intra-class sparsity predicts class-wise memorization}%
\label{sec:class-wise-memorisation}

\begin{figure*}[!t]
    \centering
    \includegraphics[width=\linewidth]{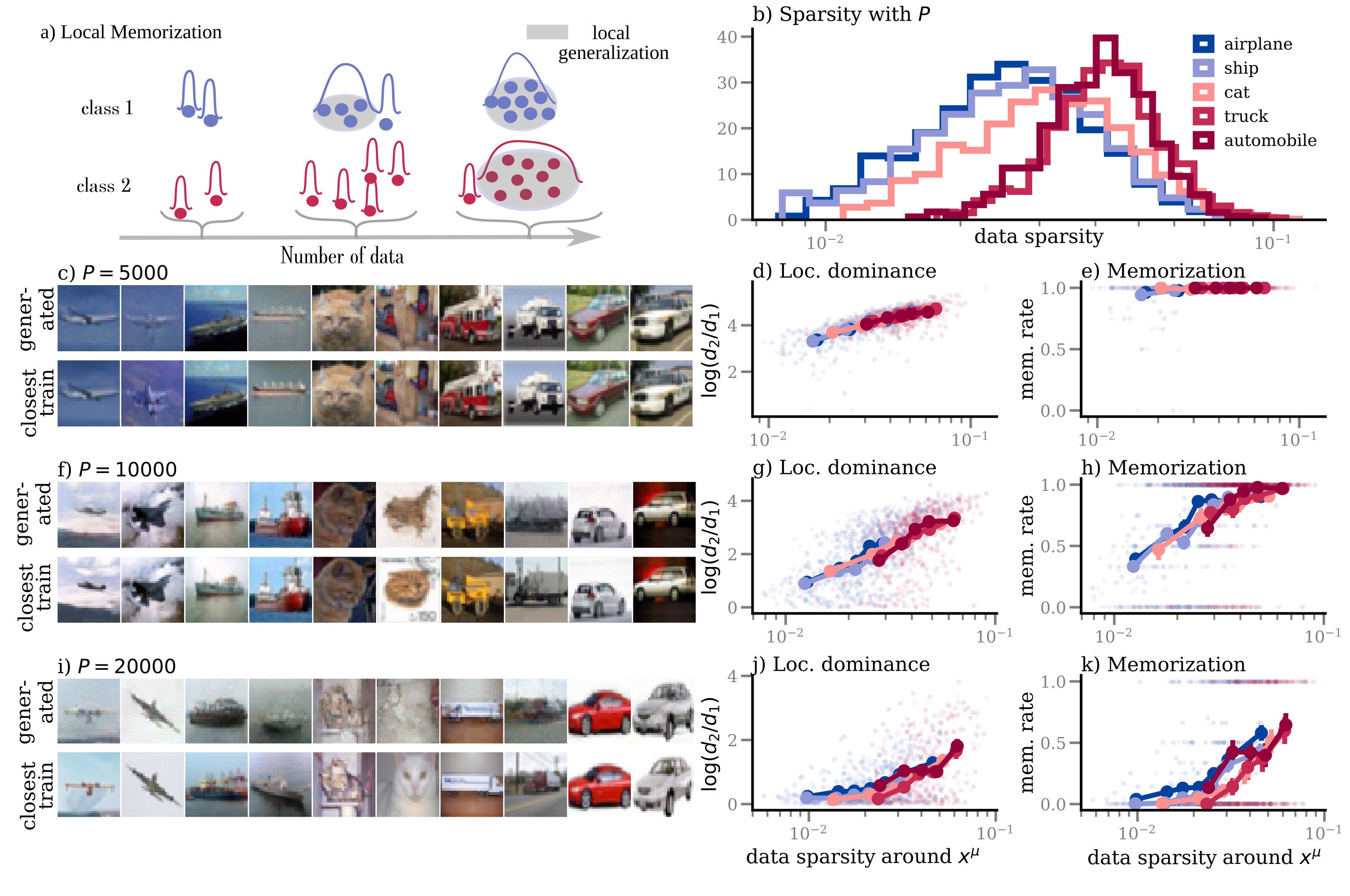}
    \caption{\textbf{Intra-class sparsity predicts class-wise memorization.}
    a) Schematic illustration of KDE for two classes with different local sparsities.
    b) Distribution of per-class sparsity around training points for different subclasses of CIFAR-10, measured from $20{,}000$ training samples.
    c) Generated samples together with their closest training example (measured by cosine similarity) for a diffusion model trained on $P=1000$ samples, sorted to contain 2 nearest neighbors for each of the four classes shown in b).
    d) Local dominance as a function of sparsity. For each generated sample $\hat{x}$, we compute distances $d_1$ and $d_2$ to its nearest and second-nearest training samples, and define dominance via their ratio. Each point corresponds to a training sample $x^\mu$, aggregating generated samples assigned to it. Scatter shows raw values, markers indicate binned averages. Different colors correspond to different classes.
    e) Memorization rate of each sample $x^\mu$, defined as the fraction of generated samples satisfying $d_1/d_2 < \tau = 1/3$.
    f) - k) report the same measures, but for diffusion models trained on larger datasets.
    "Airplane" and "ship" samples lie in denser regions and exhibit lower dominance and memorization, while "truck" and "automobile" samples are sparser and more frequently memorized, consistent with the prediction that local coverage controls class-dependent memorization.
}
    \label{fig:classwise_cifar}
\end{figure*}

We now study memorization at a coarser level, focusing on \textbf{class-dependent effects}. In a multi-class setting, the KDE picture predicts that classes with more concentrated support (i.e.\ lower local sparsity) should be memorized less than classes with higher intra-class variability. A schematic illustration is shown in \cref{fig:classwise_cifar}a.
To formalize this intuition, consider a mixture of class-conditional densities $\rho_c$, such that
\begin{equation*} 
\textstyle
    \rho(x) = \sum_c w_c \rho_c(x),
\end{equation*}
where the weights $w_c$ sum to one, and we assume that all weights are order one in $N$. We define the local in-distribution log-density for class $c$ as
\begin{equation}
    \nu_{\mathrm{in},c}(x) := \frac{1}{N} \ln p_{\mathrm{in},c}(x)
    = \frac{1}{N} \ln \int_{B_h(x)} d\rho_c(x'),
\end{equation}
and assume that $\nu_{\mathrm{in},c}(x)$ remains $\mathcal{O}(1)$ in the large-$N$ limit. Then the total in-distribution density satisfies
\begin{equation*}
    \nu_{\mathrm{in}}(x)
    = \frac{1}{N} \ln \sum_c e^{N\left(\nu_{\mathrm{in},c}(x) + N^{-1} \ln w_c\right)}
    \;\longrightarrow\;
    \max_c \nu_{\mathrm{in},c}(x),
\end{equation*}
i.e.\ it is dominated by the locally densest class. Consequently, we obtain the same scaling behavior as in lemma \ref{lemma:local_kde_transition}, in a class-dependent manner:
\begin{equation}
    z(x)\;\rightarrow\;
    \begin{cases}
        \max_c \nu_{\mathrm{in},c}(x) & \text{if } \max_c \nu_{\mathrm{in},c}(x) \geq -\alpha, \\
        -\infty & \text{otherwise},
    \end{cases}
\end{equation}
showing that memorization is governed by the class with the highest local coverage around $x$. This leads to a clear prediction: classes with higher intra-class variability (and thus lower local coverage) should be more prone to memorization.

We now demonstrate empirically that such class-dependent density differences naturally arise. As a concrete example, we isolate the memorization behavior of different sub-classes of CIFAR-10. In \cref{fig:classwise_cifar}b we show that the classes "airplane" and "ship" in CIFAR-10 have, on average, considerably lower intra-class diversity than the classes "truck" and "automobile", likely due to more homogeneous blue background colors in the former two classes. Correspondingly, we observe both higher local dominance, and higher memorization in more diverse classes. 

In the appendix, \cref{app:additional experiments}, we show that the same trends emerge for two additional experiments. 
First, when one sorts the CelebA datasets into classes according to whether images have the attribute "wearing hat" or "blond",  images where "wearing hat" is true are more diverse than those where it is false. Correspondingly, portraits featuring hats are memorized more.

Second, we train diffusion models on a mixture of MNIST \citep{deng_mnist_2012} and CIFAR-10 images. MNIST, consisting of handwritten digits, is structurally simpler and occupies a much more concentrated region of image space, whereas CIFAR-10 exhibits substantially higher variability. Treating MNIST and CIFAR-10 as two distinct classes, we observe that MNIST samples consistently exhibit lower local sparsity than CIFAR-10 samples. As predicted by the theory, this translates into reduced memorization. 
These results support the hypothesis that classes with higher intra-class diversity are memorized more strongly than classes with lower intra-class diversity. 

\section{Discussion}

\paragraph{Summary.} In this work, we test the predictive power of kernel
density estimation to explain local memorization in diffusion models. While
earlier works \citep{lucibello_exponential_2024, biroli_kernel_2024,
achilli_memorization_2025, achilli_losing_2026} focused on establishing \textit{global} phase
transitions from memorization to generalization for \textit{typical} samples from
the data distribution, we developed a \emph{local} theory of memorisation for KDE. We
showed that, in the high-dimensional limit, each point $x$ in sample space
undergoes a local transition from memorisation to generalisation as a function
of sample complexity. This transition is governed by \emph{local coverage} of
training data, which separates regions where the density is smoothly estimated
from regions dominated by isolated training samples. This leads to a coexistence
of \emph{generalization} and \emph{memorization} within the same model. Our
analysis further predicts that, in multi-class settings, the behavior at a point
$x$ is controlled by the class with the highest local coverage. Empirically, we
find consistent evidence for these predictions: memorization increases with
local sparsity and intra-class variability, and different classes exhibit
systematically different memorization behavior. Taken together, these results
show that memorization in diffusion models is not a global failure mode, but a
\emph{localized phenomenon} driven by the geometry of the data distribution. In
particular, isolated regions of the data space can remain memorized even when
the majority of generated samples are novel.

\paragraph{Limitations.}
Our analysis relies on the infinite-dimensional limit $N\rightarrow \infty$ with an exponential sample complexity $P=e^{\alpha N}$,  analogous to prior work \citep{lucibello_exponential_2024, biroli_kernel_2024}. While this enables a sharp theoretical characterization, it predicts an abrupt transition between memorization and generalization as a function of local coverage, whereas empirically we observe a more gradual dependence on local sparsity. This more gradual behavior could be the result of finite-size fluctuations as well as an extended region where the second condition in \cref{thm:W_regimes} is fulfilled, which allows for fluctuations even in the infinite limit. While we hypothesize that finite-$N$ fluctuations are at the core of this more gradual increase, disentangling these effects from fluctuations which persist in the infinite limit is an important direction for future work. Similarly, the qualitative equivalence between hard-sphere kernels and Gaussian kernels, which we exploit here to make the dependence of KDE on local coverage explicit, is expected to increase with the dimension $N$. We leave an investigation into the effects of the exact shape of the kernel to future work.

A second limitation stems from the choice of the metric space underlying the KDE approximation. Our theory assumes that diffusion models operate in the ambient space. In practice, however, models may implicitly operate in a lower-dimensional representation, for example through projection onto a data manifold \citep{achilli_capacity_2025, achilli_losing_2026} or by exploiting symmetries in the data \citep{kamb_analytic_2025}, or as latent diffusion models \cite{rombach_high-resolution_2022}. In such cases, the effective kernel should be understood as acting in this learned metric space, which depends on model architecture and may help explain the observed dependence of memorization on model capacity \citep{yoon_diffusion_2023}, which has already been shown explicitly for autoencoder architectures by \citet{zhang_generalization_2025}.

\paragraph{Outlook.}
Our results suggest that controlling memorization requires shaping the \emph{local geometry} of the data representation. In particular, learning representations that increase local coverage, by mapping data to lower-dimensional spaces or by exploiting invariance may reduce memorization but reduces model expressivity, presenting a tradeoff between increased local coverage and preserving generative performance.

Another important direction concerns training dynamics. Memorization is known to depend on training time and optimization hyperparameters \citep{bonnaire2026diffusion, favero_bigger_2025, wu_taking_2025}, and recent work suggests that memorization and generalization can emerge simultaneously during training \citep{garnier-brun_biased_2026}. Understanding how convergence to the KDE solution in the generalized and memorized regimes depends on optimization dynamics is a promising direction for future research.

\paragraph{Broader impact.} While this work is foundational, memorization in
generative models is directly related to privacy and copyright risks.
Understanding which examples are likely to be memorized may help audit and
mitigate training-data reproduction, although the same insights could
potentially inform extraction attempts. At the same time, a better understanding
of memorisation might help devise strategies to deliberately store samples in
diffusion model.

\section*{Acknowledgements}
We thank Luca Mazzucco, Fabiola Ricci and Enrico Ventura for valuable discussions. CM and SG gratefully acknowledge funding from Next Generation EU, in the context
of the National Recovery and Resilience Plan, Investment PE1 -- Project FAIR
``Future Artificial Intelligence Research’’ (CUP G53C22000440006). SG
additionally acknowledges funding from the European Research Council (ERC) for
the project ``beyond2'', ID 101166056, and from the European
Union--NextGenerationEU, in the framework of the PRIN Project SELF-MADE (code
2022E3WYTY – CUP G53D23000780001).

\newpage

\printbibliography

@InProceedings{	  george2026denoising,
  title		= {Denoising Score Matching with Random Features: Insights on
		  Diffusion Models From Precise Learning Curves},
  author	= {Anand Jerry George and Rodrigo Veiga and Nicolas Macris},
  booktitle	= {The 29th International Conference on Artificial
		  Intelligence and Statistics},
  year		= {2026},
  url		= {https://openreview.net/forum?id=ZnplHm2uRt}
}

@Article{	  maillard2026memorisation,
  title		= {Memorisation, convergence and generalisation in generative
		  models},
  author	= {Maillard, Antoine and Goldt, Sebastian},
  journal	= {arXiv preprint arXiv:2605.21402},
  year		= {2026}
}

@Article{	  wang2026random,
  title		= {A Random Matrix Theory Perspective on the Consistency of
		  Diffusion Models},
  author	= {Wang, Binxu and Zavatone-Veth, Jacob and Pehlevan,
		  Cengiz},
  journal	= {arXiv preprint arXiv:2602.02908},
  year		= {2026}
}

@InProceedings{	  zhang_generalization_2025,
  title		= {Generalization of {Diffusion} {Models} {Arises} with a
		  {Balanced} {Representation} {Space}},
  url		= {https://openreview.net/forum?id=57THeGgNAN},
  urldate	= {2026-05-26},
  author	= {Zhang, Zekai and Li, Xiao and Li, Xiang and Shi, Lianghe
		  and Wu, Meng and Tao, Molei and Qu, Qing},
  month		= oct,
  year		= {2025}
}

@Misc{		  rombach_high-resolution_2022,
  title		= {High-{Resolution} {Image} {Synthesis} with {Latent}
		  {Diffusion} {Models}},
  url		= {http://arxiv.org/abs/2112.10752},
  doi		= {10.48550/arXiv.2112.10752},
  urldate	= {2026-05-26},
  publisher	= {arXiv},
  author	= {Rombach, Robin and Blattmann, Andreas and Lorenz, Dominik
		  and Esser, Patrick and Ommer, Björn},
  month		= apr,
  year		= {2022},
  note		= {arXiv:2112.10752 [cs.CV]}
}

@Article{	  achilli_memorization_2025,
  title		= {Memorization and generalization in generative diffusion
		  under the manifold hypothesis},
  volume	= {2025},
  issn		= {1742-5468},
  url		= {https://doi.org/10.1088/1742-5468/ade136},
  doi		= {10.1088/1742-5468/ade136},
  language	= {en},
  number	= {7},
  urldate	= {2025-11-13},
  journal	= {Journal of Statistical Mechanics: Theory and Experiment},
  publisher	= {IOP Publishing},
  author	= {Achilli, Beatrice and Ambrogioni, Luca and Lucibello,
		  Carlo and Mézard, Marc and Ventura, Enrico},
  month		= jul,
  year		= {2025},
  pages		= {073401}
}

@Misc{		  garnier-brun_biased_2026,
  title		= {Biased {Generalization} in {Diffusion} {Models}},
  url		= {http://arxiv.org/abs/2603.03469},
  doi		= {10.48550/arXiv.2603.03469},
  urldate	= {2026-05-04},
  publisher	= {arXiv},
  author	= {Garnier-Brun, Jerome and Biggio, Luca and Beltrame, Davide
		  and Mézard, Marc and Saglietti, Luca},
  month		= mar,
  year		= {2026},
  note		= {arXiv:2603.03469 [cs]}
}

@InProceedings{	  kamb_analytic_2025,
  title		= {An analytic theory of creativity in convolutional
		  diffusion models},
  url		= {https://openreview.net/forum?id=ilpL2qACla},
  language	= {en},
  urldate	= {2026-05-04},
  author	= {Kamb, Mason and Ganguli, Surya},
  month		= jun,
  year		= {2025}
}

@Misc{		  li_good_2024,
  title		= {A {Good} {Score} {Does} not {Lead} to {A} {Good}
		  {Generative} {Model}},
  url		= {http://arxiv.org/abs/2401.04856},
  doi		= {10.48550/arXiv.2401.04856},
  language	= {en},
  urldate	= {2026-05-04},
  publisher	= {arXiv},
  author	= {Li, Sixu and Chen, Shi and Li, Qin},
  month		= jan,
  year		= {2024},
  note		= {arXiv:2401.04856 [cs]}
}

@InProceedings{	  wu_taking_2025,
  title		= {Taking a {Big} {Step}: {Large} {Learning} {Rates} in
		  {Denoising} {Score} {Matching} {Prevent} {Memorization}},
  issn		= {2640-3498},
  shorttitle	= {Taking a {Big} {Step}},
  url		= {https://proceedings.mlr.press/v291/wu25a.html},
  language	= {en},
  urldate	= {2026-05-04},
  booktitle	= {Proceedings of {Thirty} {Eighth} {Conference} on
		  {Learning} {Theory}},
  publisher	= {PMLR},
  author	= {Wu, Yu-Han and Marion, Pierre and Biau, Gérard and Boyer,
		  Claire},
  month		= jul,
  year		= {2025},
  pages		= {5718--5756}
}

@InProceedings{	  pidstrigach_score-based_2022,
  title		= {Score-{Based} {Generative} {Models} {Detect} {Manifolds}},
  url		= {https://openreview.net/forum?id=AiNrnIrDfD9},
  language	= {en},
  urldate	= {2026-04-30},
  author	= {Pidstrigach, Jakiw},
  month		= oct,
  year		= {2022}
}

@Article{	  wen_detecting_2024,
  title		= {{DETECTING}, {EXPLAINING}, {AND} {MITIGATING} {MEMO}-
		  {RIZATION} {IN} {DIFFUSION} {MODELS}},
  language	= {en},
  author	= {Wen, Yuxin and Liu, Yuchen and Chen, Chen and Lyu,
		  Lingjuan},
  year		= {2024}
}

@Misc{		  jeon_understanding_2025,
  title		= {Understanding and {Mitigating} {Memorization} in
		  {Generative} {Models} via {Sharpness} of {Probability}
		  {Landscapes}},
  url		= {http://arxiv.org/abs/2412.04140},
  doi		= {10.48550/arXiv.2412.04140},
  language	= {en},
  urldate	= {2026-04-26},
  publisher	= {arXiv},
  author	= {Jeon, Dongjae and Kim, Dueun and No, Albert},
  month		= aug,
  year		= {2025},
  note		= {arXiv:2412.04140 [cs]}
}

@Misc{		  fang_closer_2025,
  title		= {A {Closer} {Look} on {Memorization} in {Tabular}
		  {Diffusion} {Model}: {A} {Data}-{Centric} {Perspective}},
  shorttitle	= {A {Closer} {Look} on {Memorization} in {Tabular}
		  {Diffusion} {Model}},
  url		= {http://arxiv.org/abs/2505.22322},
  doi		= {10.48550/arXiv.2505.22322},
  language	= {en},
  urldate	= {2026-04-26},
  publisher	= {arXiv},
  author	= {Fang, Zhengyu and Jiang, Zhimeng and Chen, Huiyuan and
		  Zhang, Xiaoge and Tang, Kaiyu and Li, Xiao and Li, Jing},
  month		= aug,
  year		= {2025},
  note		= {arXiv:2505.22322 [cs]}
}

@Misc{		  kim_how_2025,
  title		= {How {Diffusion} {Models} {Memorize}},
  url		= {http://arxiv.org/abs/2509.25705},
  doi		= {10.48550/arXiv.2509.25705},
  language	= {en},
  urldate	= {2026-04-26},
  publisher	= {arXiv},
  author	= {Kim, Juyeop and Kim, Songkuk and Lee, Jong-Seok},
  month		= sep,
  year		= {2025},
  note		= {arXiv:2509.25705 [cs]}
}

@Misc{		  gu_memorization_2025,
  title		= {On {Memorization} in {Diffusion} {Models}},
  url		= {http://arxiv.org/abs/2310.02664},
  doi		= {10.48550/arXiv.2310.02664},
  language	= {en},
  urldate	= {2026-04-26},
  publisher	= {arXiv},
  author	= {Gu, Xiangming and Du, Chao and Pang, Tianyu and Li,
		  Chongxuan and Lin, Min and Wang, Ye},
  month		= feb,
  year		= {2025},
  note		= {arXiv:2310.02664 [cs]}
}

@InProceedings{	  yoon_diffusion_2023,
  title		= {Diffusion {Probabilistic} {Models} {Generalize} when
		  {They} {Fail} to {Memorize}},
  url		= {https://openreview.net/forum?id=shciCbSk9h#all},
  language	= {en},
  urldate	= {2026-04-26},
  author	= {Yoon, TaeHo and Choi, Joo Young and Kwon, Sehyun and Ryu,
		  Ernest K.},
  month		= jul,
  year		= {2023}
}

@Misc{		  carlini_extracting_2023,
  title		= {Extracting {Training} {Data} from {Diffusion} {Models}},
  url		= {http://arxiv.org/abs/2301.13188},
  doi		= {10.48550/arXiv.2301.13188},
  urldate	= {2026-04-26},
  publisher	= {arXiv},
  author	= {Carlini, Nicholas and Hayes, Jamie and Nasr, Milad and
		  Jagielski, Matthew and Sehwag, Vikash and Tramèr, Florian
		  and Balle, Borja and Ippolito, Daphne and Wallace, Eric},
  year		= {2023},
  note		= {arXiv:2301.13188}
}

@Misc{		  ross_geometric_2025,
  title		= {A {Geometric} {Framework} for {Understanding}
		  {Memorization} in {Generative} {Models}},
  url		= {http://arxiv.org/abs/2411.00113},
  doi		= {10.48550/arXiv.2411.00113},
  urldate	= {2026-04-23},
  publisher	= {arXiv},
  author	= {Ross, Brendan Leigh and Kamkari, Hamidreza and Wu, Tongzi
		  and Hosseinzadeh, Rasa and Liu, Zhaoyan and Stein, George
		  and Cresswell, Jesse C. and Loaiza-Ganem, Gabriel},
  month		= mar,
  year		= {2025},
  note		= {arXiv:2411.00113 [stat]}
}

@Misc{		  achilli_losing_2026,
  title		= {Losing dimensions: {Geometric} memorization in generative
		  diffusion},
  shorttitle	= {Losing dimensions},
  url		= {http://arxiv.org/abs/2410.08727},
  doi		= {10.48550/arXiv.2410.08727},
  language	= {en},
  urldate	= {2026-04-21},
  publisher	= {arXiv},
  author	= {Achilli, Beatrice and Ventura, Enrico and Silvestri,
		  Gianluigi and Pham, Bao and Raya, Gabriel and Krotov,
		  Dmitry and Lucibello, Carlo and Ambrogioni, Luca},
  month		= mar,
  year		= {2026},
  note		= {arXiv:2410.08727 [stat]}
}

@Article{	  ambrogioni_search_2024,
  title		= {In {Search} of {Dispersed} {Memories}: {Generative}
		  {Diffusion} {Models} {Are} {Associative} {Memory}
		  {Networks}},
  volume	= {26},
  copyright	= {http://creativecommons.org/licenses/by/3.0/},
  issn		= {1099-4300},
  shorttitle	= {In {Search} of {Dispersed} {Memories}},
  url		= {https://www.mdpi.com/1099-4300/26/5/381},
  doi		= {10.3390/e26050381},
  number	= {5},
  urldate	= {2025-11-04},
  journal	= {Entropy},
  publisher	= {Multidisciplinary Digital Publishing Institute},
  author	= {Ambrogioni, Luca},
  year		= {2024},
  pages		= {381}
}

@Misc{		  achilli_capacity_2025,
  title		= {The {Capacity} of {Modern} {Hopfield} {Networks} under the
		  {Data} {Manifold} {Hypothesis}},
  url		= {http://arxiv.org/abs/2503.09518},
  doi		= {10.48550/arXiv.2503.09518},
  language	= {en},
  urldate	= {2025-10-27},
  publisher	= {arXiv},
  author	= {Achilli, Beatrice and Ambrogioni, Luca and Lucibello,
		  Carlo and Mézard, Marc and Ventura, Enrico},
  month		= mar,
  year		= {2025},
  note		= {arXiv:2503.09518 [cond-mat]}
}

@InProceedings{	  krizhevsky_learning_2009,
  title		= {Learning {Multiple} {Layers} of {Features} from {Tiny}
		  {Images}},
  url		= {https://www.semanticscholar.org/paper/Learning-Multiple-Layers-of-Features-from-Tiny-Krizhevsky/5d90f06bb70a0a3dced62413346235c02b1aa086},
  urldate	= {2025-10-09},
  author	= {Krizhevsky, A.},
  year		= {2009}
}

@Misc{		  favero_bigger_2025,
  title		= {Bigger {Isn}'t {Always} {Memorizing}: {Early} {Stopping}
		  {Overparameterized} {Diffusion} {Models}},
  shorttitle	= {Bigger {Isn}'t {Always} {Memorizing}},
  url		= {http://arxiv.org/abs/2505.16959},
  doi		= {10.48550/arXiv.2505.16959},
  urldate	= {2025-09-18},
  publisher	= {arXiv},
  author	= {Favero, Alessandro and Sclocchi, Antonio and Wyart,
		  Matthieu},
  month		= sep,
  year		= {2025},
  note		= {arXiv:2505.16959 [cs]}
}

@Article{	  bonnaire2026diffusion,
  title		= {Why diffusion models don’t memorize: The role of
		  implicit dynamical regularization in training},
  author	= {Bonnaire, Tony and Urfin, Rapha{\"e}l and Biroli, Giulio
		  and M{\'e}zard, Marc},
  journal	= {Advances in Neural Information Processing Systems},
  volume	= {38},
  pages		= {141266--141286},
  year		= {2026}
}

@Article{	  lucibello_exponential_2024,
  title		= {Exponential {Capacity} of {Dense} {Associative}
		  {Memories}},
  volume	= {132},
  url		= {https://link.aps.org/doi/10.1103/PhysRevLett.132.077301},
  doi		= {10.1103/PhysRevLett.132.077301},
  number	= {7},
  urldate	= {2025-08-18},
  journal	= {Physical Review Letters},
  publisher	= {American Physical Society},
  author	= {Lucibello, Carlo and Mézard, Marc},
  month		= feb,
  year		= {2024},
  pages		= {077301}
}

@Article{	  dotsenko_replica_2011,
  title		= {Replica solution of the random energy model},
  volume	= {95},
  issn		= {0295-5075, 1286-4854},
  url		= {https://iopscience.iop.org/article/10.1209/0295-5075/95/50006},
  doi		= {10.1209/0295-5075/95/50006},
  language	= {en},
  number	= {5},
  urldate	= {2025-08-02},
  journal	= {EPL (Europhysics Letters)},
  author	= {Dotsenko, V.},
  month		= sep,
  year		= {2011},
  pages		= {50006}
}

@Misc{		  biroli_kernel_2024,
  title		= {Kernel {Density} {Estimators} in {Large} {Dimensions}},
  url		= {http://arxiv.org/abs/2408.05807},
  doi		= {10.48550/arXiv.2408.05807},
  urldate	= {2025-06-27},
  publisher	= {arXiv},
  author	= {Biroli, Giulio and Mézard, Marc},
  month		= oct,
  year		= {2024},
  note		= {arXiv:2408.05807 [cs]}
}

@Article{	  biroli_dynamical_2024,
  title		= {Dynamical regimes of diffusion models},
  volume	= {15},
  copyright	= {2024 The Author(s)},
  issn		= {2041-1723},
  url		= {https://www.nature.com/articles/s41467-024-54281-3},
  doi		= {10.1038/s41467-024-54281-3},
  language	= {en},
  number	= {1},
  urldate	= {2025-06-27},
  journal	= {Nature Communications},
  publisher	= {Nature Publishing Group},
  author	= {Biroli, Giulio and Bonnaire, Tony and de Bortoli, Valentin
		  and Mézard, Marc},
  month		= nov,
  year		= {2024},
  pages		= {9957}
}

@Misc{		  ronneberger_u-net_2015,
  title		= {U-{Net}: {Convolutional} {Networks} for {Biomedical}
		  {Image} {Segmentation}},
  shorttitle	= {U-{Net}},
  url		= {http://arxiv.org/abs/1505.04597},
  doi		= {10.48550/arXiv.1505.04597},
  urldate	= {2025-05-15},
  publisher	= {arXiv},
  author	= {Ronneberger, Olaf and Fischer, Philipp and Brox, Thomas},
  month		= may,
  year		= {2015},
  note		= {arXiv:1505.04597 [cs]}
}

@Misc{		  george_denoising_2025,
  title		= {Denoising {Score} {Matching} with {Random} {Features}:
		  {Insights} on {Diffusion} {Models} from {Precise}
		  {Learning} {Curves}},
  shorttitle	= {Denoising {Score} {Matching} with {Random} {Features}},
  url		= {http://arxiv.org/abs/2502.00336},
  doi		= {10.48550/arXiv.2502.00336},
  urldate	= {2025-05-13},
  publisher	= {arXiv},
  author	= {George, Anand Jerry and Veiga, Rodrigo and Macris,
		  Nicolas},
  month		= feb,
  year		= {2025},
  note		= {arXiv:2502.00336 [cs]}
}

@Article{	  deng_mnist_2012,
  title		= {The {MNIST} {Database} of {Handwritten} {Digit} {Images}
		  for {Machine} {Learning} {Research} [{Best} of the {Web}]},
  volume	= {29},
  issn		= {1558-0792},
  url		= {https://ieeexplore.ieee.org/document/6296535},
  doi		= {10.1109/MSP.2012.2211477},
  number	= {6},
  urldate	= {2025-05-10},
  journal	= {IEEE Signal Processing Magazine},
  author	= {Deng, Li},
  month		= nov,
  year		= {2012},
  pages		= {141--142}
}

@InProceedings{	  liu_deep_2015,
  title		= {Deep {Learning} {Face} {Attributes} in the {Wild}},
  isbn		= {978-1-4673-8391-2},
  issn		= {2380-7504},
  url		= {https://www.computer.org/csdl/proceedings-article/iccv/2015/8391d730/12OmNzGlRCR},
  doi		= {10.1109/ICCV.2015.425},
  language	= {English},
  urldate	= {2025-05-10},
  publisher	= {IEEE Computer Society},
  author	= {Liu, Ziwei and Luo, Ping and Wang, Xiaogang and Tang,
		  Xiaoou},
  month		= dec,
  year		= {2015},
  pages		= {3730--3738}
}

@InProceedings{	  somepalli_understanding_2023,
  title		= {Understanding and {Mitigating} {Copying} in {Diffusion}
		  {Models}},
  url		= {https://openreview.net/forum?id=HtMXRGbUMt},
  language	= {en},
  urldate	= {2025-05-08},
  author	= {Somepalli, Gowthami and Singla, Vasu and Goldblum, Micah
		  and Geiping, Jonas and Goldstein, Tom},
  month		= nov,
  year		= {2023}
}

@Misc{		  somepalli_diffusion_2022,
  title		= {Diffusion {Art} or {Digital} {Forgery}? {Investigating}
		  {Data} {Replication} in {Diffusion} {Models}},
  shorttitle	= {Diffusion {Art} or {Digital} {Forgery}?},
  url		= {http://arxiv.org/abs/2212.03860},
  doi		= {10.48550/arXiv.2212.03860},
  urldate	= {2025-05-08},
  publisher	= {arXiv},
  author	= {Somepalli, Gowthami and Singla, Vasu and Goldblum, Micah
		  and Geiping, Jonas and Goldstein, Tom},
  month		= dec,
  year		= {2022},
  note		= {arXiv:2212.03860 [cs]}
}

@InProceedings{	  pham_memorization_2024,
  title		= {Memorization to {Generalization}: {The} {Emergence} of
		  {Diffusion} {Models} from {Associative} {Memory}},
  shorttitle	= {Memorization to {Generalization}},
  url		= {https://openreview.net/forum?id=zVMMaVy2BY},
  language	= {en},
  urldate	= {2025-05-08},
  author	= {Pham, Bao and Raya, Gabriel and Negri, Matteo and Zaki,
		  Mohammed J. and Ambrogioni, Luca and Krotov, Dmitry},
  month		= nov,
  year		= {2024}
}

@InProceedings{	  kadkhodaie2024generalization,
  author	= {Kadkhodaie, Zahra and Guth, Florentin and Simoncelli, Eero
		  and Mallat, St\'{e}phane},
  booktitle	= {International Conference on Learning Representations},
  editor	= {B. Kim and Y. Yue and S. Chaudhuri and K. Fragkiadaki and
		  M. Khan and Y. Sun},
  pages		= {46543--46567},
  title		= {Generalization in diffusion models arises from
		  geometry-adaptive harmonic representations},
  url		= {https://proceedings.iclr.cc/paper_files/paper/2024/file/cbaf319a4712385b5ba8a414808b5713-Paper-Conference.pdf},
  volume	= {2024},
  year		= {2024}
}

\newpage
\appendix
\onecolumn
\numberwithin{equation}{section}
\numberwithin{figure}{section}

\section{Details on Theoretical results} 
\label{app: Derivations}
In this appendix, we give a detailed proofs of the results presented in \cref{sec:theory}. 

\subsection{Convergence of the exponential transform}
\label{app:convergence}
To establish the behavior of the KDE, we characterize the cumulative distribution function (CDF) of the log-density $z$ through the exponential transform \cref{eq:W_N_definition}. At the base of \cref{thm:W_regimes} therefore lies the convergence of the exponential transform to the CDF, which is provided by the following Lemma. 
\begin{lemma}[Convergence of the exponential transform to the CDF.]
\label{lemma:convergence}
Let $N \in \mathbb{N}, N>0, x_N \in R^N$ and $F_N(y) = \Pr (z_N(x_N) \le y)$ be the cumulative distribution function of $z_N(x_N)$. Then for $\overline{W_N(y,x_N)}$ defined as in \cref{eq:W_N_definition} and $\delta > 0$, we find that 
\begin{equation}
    e^{-e^{-N\delta}}F_N(y-\delta)
    \le
    \overline{ W_N(y, x_N)} 
    \le
    F_N(y+\delta)+e^{-e^{N\delta}} \Pr (z_N\ge y+\delta)..
\end{equation}
Consequently, if the limit $F_N\to F_{\mathbf{x}}$  exists and if $F_{\mathbf{x}}$ is continuous in $y$, then
\[
\overline{ W_N(y,x_N) }\to F_{\mathbf{x}}(y).
\]
\end{lemma}

Proof. For notational simplicity we suppress the
dependence on \(x\) and write
\begin{equation*}
    z_N := z_N(x),
    \qquad
    W_N(y) := \exp\{-\exp[-N(y-z_N)]\}.
\end{equation*}
We first show the lower bound. If $z_N\le y-\delta$, then
\begin{equation*}
     W_N(y) = \exp\{-\exp[-N(y-z_N)]\}
    \ge
    \exp\{-e^{-N\delta}\}.
\end{equation*}
In other words, using the indicator function $\mathbf{1}$, we know that 
\begin{equation*}
    W_N(y)
    \ge
    e^{-e^{-N\delta}}\mathbf 1\{z_N\le y-\delta\}.
\end{equation*}
Taking expectations gives
\begin{equation*}
    \overline{W_N(y)}
    \ge
    e^{-e^{-N\delta}}\mathbb P(z_N\le y-\delta).
\end{equation*}
We now take care of the upper bound. Since \(0\le W_N(y)\le 1\), we know that 
\begin{align*}
    W_N(y)
    & \le 
    \mathbf 1\{z_N<y+\delta\}
    +
    W_N(y) \mathbf 1\{z_N\ge y+\delta\} \\
    & \le
    \mathbf 1\{z_N<y+\delta\}
    +
    e^{-e^{N\delta}}\mathbf 1\{z_N\ge y+\delta\}.
\end{align*}
Taking the average yields 
\begin{equation*}
    \overline{ W_N(y)} 
    \le
    \mathbb P(z_N<y+\delta)+e^{-e^{N\delta}} \mathbb P(z_N\ge y+\delta).
\end{equation*}
Combining the two bounds yields Lemma \ref{lemma:convergence}. \qed

\subsection{Average over draws of samples \cref{thm:W_regimes}}
We now proceed with the proof of \cref{thm:W_regimes}. Let us first keep $N \in \mathbb{N}, N>0, x_N \in R^N$ fixed. 
Our starting point is \cref{eq:W_N_definition}. Recall that the notation $\overline{f(z)}$ denotes the average over the data set, meaning that when we insert the definition of $z(x)$ and explicitly state the average, we find 
\begin{align*}
     \overline{W_{N}(y,x_N)}
    =\int \prod_{\mu} d\rho(x^{\mu}) \exp\left(-e^{-N\left(y+\alpha\right)}
        \sum_{\mu} K(x-x^{\mu})\right) \\
    = \left\{ 
    \int d\rho(\tilde{x}) \exp\left[-e^{-N\left(y+\alpha\right)} 
         K\left(x-\tilde{x}\right) \right]  \right\}^P
\end{align*}
which directly yields \cref{eq:W_N_averaged}. Taking the limit then reveals the three scaling regimes
\begin{equation}
    \overline{W_{N}(y,x_N)} \rightarrow
    \begin{cases}
        0 & \text{if } i_{\mathbf{x}}(y)=\infty,\\
        \exp[-i_{\mathbf{x}}(y)] & \text{if } 0<i_{\mathbf{x}}(y)<\infty,\\
        1 & \text{if } i_{\mathbf{x}}(y)=0.
    \end{cases}
\end{equation}

Using, lemma \ref{lemma:convergence}, we then find that the limit of the exponential transform yields the limiting CDF. \qed 

\subsection{Hard spherical kernels}
We now turn to the proof of lemma \ref{lemma:local_kde_transition}. Using the definition of $p_{\text{in}}$, we find that the integral $I_N$ defined in \cref{eq:I_N_kernel_statistic} decomposes into two areas: those where $K=0$ and those where $K=1/V_R$. Using this distinction, we then find
\begin{equation}
    I_{N}\left(y,x_N\right)= p_{\text{in},N}(x_N)\left(1-\exp\left(-\exp\left\{ -N\left(y+\alpha + \frac{1}{N}  \ln V_R \right)\right\} \right)\right)
\end{equation}
Since, we have fixed $R$ such that $ \frac{1}{N}  \ln V_R \rightarrow 0 $, the latter term in the exponential vanishes in the high-dimensional limit. To apply \cref{thm:W_regimes}, we use the assumption that $\nu_{\text{in}, \mathbf{x}}$ exists and compute the limit \cref{eq:i_x} in two distinct regimes: $y+\alpha >0$ and $y+\alpha <0$. 

In the first regime, $y+\alpha=\delta>0$, we find 
\begin{align*}
    i_{\mathbf{x}} = \lim_{N\rightarrow\infty}
    e^{N (\alpha +\nu_{\text{in},N}(x_N) )} 
    \left(1-e^{-e^{-N\delta}} \right) 
    = \begin{cases}
        0 & \text{if } y- \nu_{\text{in},\mathbf{x}} >0 \\
        \infty & \text{if } y- \nu_{\text{in},\mathbf{x}} <0
    \end{cases}\,.
\end{align*}

In the second regime, $y+\alpha=-\delta<0$, we find 
\begin{align*}
    i_{\mathbf{x}} = \lim_{N\rightarrow\infty}
    e^{N (\alpha +\nu_{\text{in},N}(x_N) )} 
    \left(1-e^{-e^{N\delta}} \right) 
    = \begin{cases}
        0 & \text{if } \alpha +\nu_{\text{in},\mathbf{x}}< 0 \\
        \infty & \text{if } \alpha +\nu_{\text{in},\mathbf{x}} >0
    \end{cases}\,.
\end{align*}

Applying \cref{thm:W_regimes} to both regimes, we find 
\begin{equation*}
    F_{\mathbf{x}}(y) = \begin{cases}
\Theta\left(y-\nu_{\text{in}, \mathbf{x}}\right) & y+\alpha > 0\\
\Theta\left(-\left(\nu_{\text{in}, \mathbf{x}}+\alpha\right)\right) & y+\alpha < 0
\end{cases}
\end{equation*}
which yields the probability that $z\leq y$ in the limit, thanks to lemma \ref{lemma:convergence}. The first line shows that when $y$ is larger than $-\alpha$, the CDF becomes a step function with a jump at $y=\nu_{\text{in},\mathbf{x}}$. 

We split the second line into two scenarios,  $\nu_{\text{in},\mathbf{x}}+\alpha>0$ and $\nu_{\text{in},\mathbf{x}} +\alpha <0$
In the first, scenario $\nu_{\text{in},\mathbf{x}}>-\alpha >y$, hence this implies $y<\nu_{\text{in}}(x)$, and the limiting CDF is zero in $y$. 
On then other hand, in the second scenario, when $\nu_{\text{in},\mathbf{x}}<-\alpha$, then the CDF is one for any finite value of $y$. This means that $z$ is smaller than any finite value of $y$ with probability one.  
Taken together, the first and the second line imply that the CDF has a jump at $y=\nu_{\text{in},\mathbf{x}}$ when $\nu_{\text{in},\mathbf{x}} >-\alpha$. The second scenario in the second line implies that when $\nu_{\text{in},\mathbf{x}} <-\alpha$, all draws of $z$ are smaller than any finite $y$, hence $z\rightarrow -\infty$.

Consequently, we find that the cumulative distribution has the shape of a step function. Hence the value of $z_N(x_N)$ concentrates on the "jump location", given by 
\begin{equation}
    \begin{cases}
\nu_{\text{in}}(x) & \text{if }\nu_{\text{in}}(x)\geq-\alpha\\
-\infty & \text{if }\nu_{\text{in}}(x)<-\alpha
\end{cases}
\end{equation}
which finally yields lemma \ref{lemma:local_kde_transition}. \qed

\subsection{Relation to Gumbel-type extreme value statistics}
\label{app:gumbel_relation}

In this appendix we clarify the relation between the exponential transform used in
Eq.~(10) and Gumbel-type extreme value statistics. Let us first clarify what we mean by Gumbel-type extreme value statistics: for independent random variables \(u_1,\dots,u_P\), one has
\begin{equation}
    \mathbb{P}\!\left(\max_{\mu}u_\mu<y\right)
    =
    \left[
        1-\mathbb{P}(u_\mu\ge y)
    \right]^P.
\end{equation}
which is the standard structure associated with Gumbel-type extreme value statistics and, identifying $I_{\circ}$ with $\mathbb{P}(u_\mu\ge y)$, has strong similarity with \cref{eq:W_N_averaged}. 

We will first show that letting $N\rightarrow \infty$ at $P<\infty$ fixed, $W_N$ has a natural extreme-value interpretation. However, this does not mean that the calculation replaces the KDE by its largest term, because in the main result, we let $P\rightarrow\infty $ with $\frac{1}{N}\ln P=\alpha$ fixed. In this case, $W_N$ is a soft probe of the cumulative distribution function of the full log-KDE. 

For the sake of comparison, however, let us now keep $P<\infty$ and all samples fixed. We then find that $W_N$ can be rewritten as
\begin{align}
    W_N(y,x)
    &=
    \prod_{\mu=1}^{P}
    \exp\left\{
        -K(x-x^{\mu})e^{-N(y+\alpha)}
    \right\}.
    \label{eq:gumbel_factorized_W}
\end{align}
Now define the single-sample contribution
$
    u_\mu(x)
    :=
    \frac{1}{N}\ln K(x-x^{\mu})-\alpha .
$
Then each factor in Eq.~\eqref{eq:gumbel_factorized_W} can be written as
\begin{equation}
    \exp\left\{
        -K_\mu(x)e^{-N(y+\alpha)}
    \right\}
    =
    \exp\left\{
        -\exp[N(u_\mu(x)-y)]
    \right\}.
\end{equation}
If we were to fix $P$ and all \(u_\mu(x)\), and let $N\rightarrow\infty$ this converges to a hard threshold:
\begin{equation}
    \exp\left\{
        -\exp[N(u_\mu(x)-y)]
    \right\}
    \xrightarrow[N\to\infty]{}
    \Theta( y - u_\mu(x)).
\end{equation}
Consequently, if this hard-threshold limit is taken termwise before accounting
for the exponentially many samples, Eq.~\eqref{eq:gumbel_factorized_W} becomes
\begin{equation}
    W_N(y,x)
    \approx
    \prod_{\mu=1}^{P}
    \Theta(y-u_\mu(x))
    =
    \Theta \left( y-\max_{\mu}u_\mu(x) \right).
\end{equation}
This is the origin of the formal similarity with Gumbel's law: the product form
resembles the cumulative distribution function of a maximum. In regimes where the KDE is dominated by a single
large kernel contribution, \(z_N(x)\) is well approximated by
\begin{equation}
    \widetilde z_N(x)
    :=
    \max_{\mu}
    \frac{1}{N}\log\frac{K_\mu(x)}{P}
    =
    \max_{\mu}u_\mu(x).
\end{equation}
In such max-dominated regimes, the above Gumbel-type interpretation directly
describes the statistics of \(z_N(x)\).

However, to obtain a generalization regime, we must send $P$ to infinity too. Consequently, in the generalization regime, exponentially many samples may contribute to the KDE at any point. Then the sum cannot be replaced by its largest term.
This distinction can be seen explicitly for the hard-sphere kernel. Let \(m_N(x)\) be the number of training samples inside the corresponding ball around \(x\) with volume $V_R=1$. Then
\begin{equation}
    Z_N(x)
    =
    \frac{m_N(x)}{P},
\qquad \Rightarrow \qquad
    z_N(x)
    =
    \frac{1}{N}\log m_N(x)
    -\alpha.
\end{equation}
If the local mass of the ball satisfies $\ln p_{\mathrm{in}}(x) =
    N\nu_{\mathrm{in}}(x)+o(N),$
then the typical number of samples in the ball is
\begin{equation}
    m_N(x)
    \sim
    Pp_{\mathrm{in}}(x)
    =
    e^{N(\alpha+\nu_{\mathrm{in}}(x))+o(N)}
\end{equation}
whenever \(\alpha+\nu_{\mathrm{in}}(x)>0\). In this regime,
\begin{equation}
    z_N(x)
    \to
    \nu_{\mathrm{in}}(x),
\end{equation}
up to subexponential prefactors. By contrast, the largest single contribution is
only
\begin{equation}
    \widetilde z_N(x)
    =
    -\alpha 
\end{equation}
provided that at least one sample lies inside the ball. Thus, whenever
\(\nu_{\mathrm{in}}(x)>-\alpha\), the full KDE exponent
\(z_N(x)\) differs from the maximum contribution \(\widetilde z_N(x)\).

The Gumbel-like structure of Eq.~(11) should therefore be understood as follows:
the exponential transform is a soft thresholding device which, after
factorization over samples, resembles an extreme-value observable. If one
hardens this threshold before accounting for the exponentially many samples, one
obtains a maximum-probe interpretation. Keeping the transform soft through the
large-\(N\) calculation, however, retains the collective contribution of
exponentially many moderate terms and therefore probes the full log-KDE
\(z_N(x)\).

\section{Additional Experiments}
\label{app:additional experiments}

\subsection{Local sparsity and dominance}
\label{app:celebA}

\begin{figure*}[!htbp]
    \centering
    \includegraphics[width=\linewidth]{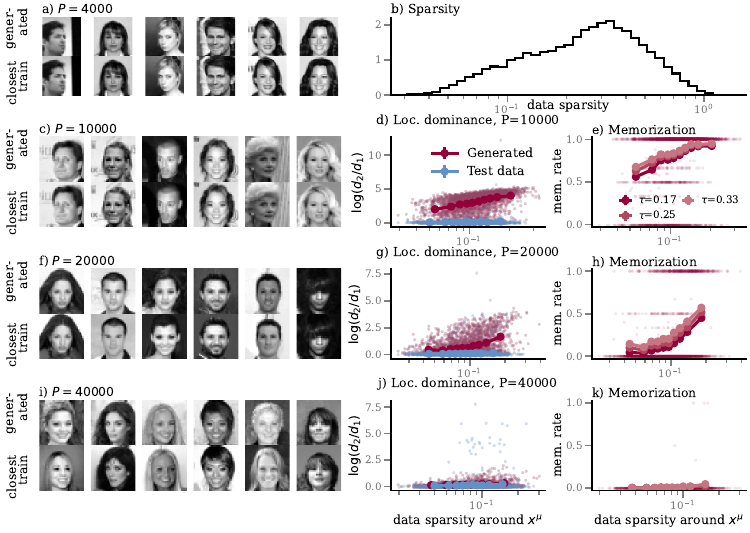}
    \caption{\textbf{Local memorization in CelebA data.} 
    a) Sketch of kernel density approximation and local memorization phenomenon. 
    b) Distribution of data sparsity for $40{,}000$ samples from CelebA.
    c) Generated samples + closest training image (measured by cosine similarity) for diffusion model trained on $P=4000$ training examples. 
    d) Local dominance as a function of sparsity. For each generated sample $\hat{x}$, we compute distances $d_1$ and $d_2$ to its nearest and second-nearest training samples, and define dominance via their ratio. Each point corresponds to a training sample $x^\mu$, aggregating generated samples assigned to it. Scatter shows raw values, markers indicate binned averages. Blue points show the same quantity computed using test data, providing a baseline without memorization.
    e) Memorization rate per training sample $x^\mu$, defined as the fraction of generated samples satisfying $d_1/d_2 < \tau$. Scatter shows raw values for $\tau=0.1$, markers indicate binned averages across thresholds.
    f)--k) Same measurements for models trained on larger datasets. As the number of training samples increases, both dominance and memorization decrease, and their dependence on sparsity weakens.
    Overall, sparse regions exhibit strong single-sample dominance and higher memorization, while dense regions promote interpolation across multiple training points.  }
    \label{fig:celeba_local_memo}
\end{figure*}

In \cref{fig:celeba_local_memo}, we report the analogous experiment to \cref{fig:fig1} for CelebA data, which we downsample to $32\times32$ greyscale pixels. We find that both local dominance and memorization increase with local sparsity. In comparison to CIFAR-10, we observe that the CelebA dataset appears to have fewer samples in very low density regions ( compare logarithmic scale of \cref{fig:celeba_local_memo} d) to \cref{fig:fig1}d). 

\subsection{Attribute-dependent Memorization, CelebA}
\begin{figure*}[!htbp]
    \centering
    \includegraphics[width=\linewidth]{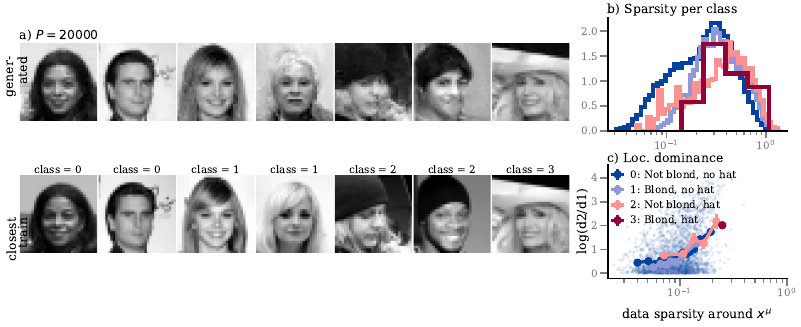}
    \caption{\textbf{Attribute-conditioned memorization in CelebA.} a) Generated samples + closest training image (measured by cosine similarity) for diffusion model trained on $P=20000$ training examples, sorted by classes. b) Distribution of data sparsity for $40000$ samples from CelebA.  d) Local dominance of training sample $x^{\mu}$ against local data sparsity around $x^{\mu}$ for a diffusion model trained on $P=20000$ training examples from CelebA.}
    \label{fig:class_conditioned_celebA}
\end{figure*}
The CelebA dataset consists of celebrity portraits that are annotated with "attributes" such as hair color or accessories. We construct four classes from these attributes, conditioning on "blond" and "not blond" as well as "wearing hat" and the opposite. Again, we find that local dominance (and thus memorization) increases with data sparsity, see \cref{fig:class_conditioned_celebA}. Samples from classes where "wearing hat" is true are typically more diverse. Therefore this class has a higher average sparsity per class,and is therefore more likely to be memorized. These classes are not balanced: approximately one sixth of images has attribute "blond" and approximately 5 \% of images have attribute "wearing hat". Consequently the diffusion model generates fewer samples that are memorized data points with these attributes, and even fewer samples are closest to training data where both attributes are true. This is reflected in the lower bin resolution in \cref{fig:class_conditioned_celebA}b) and fewer such points appearing in \cref{fig:class_conditioned_celebA} c).

\subsection{Class-dependent Memorization, CIFAR-10+MNIST}

\begin{figure*}[!htbp]
    \centering
    \includegraphics[width=\linewidth]{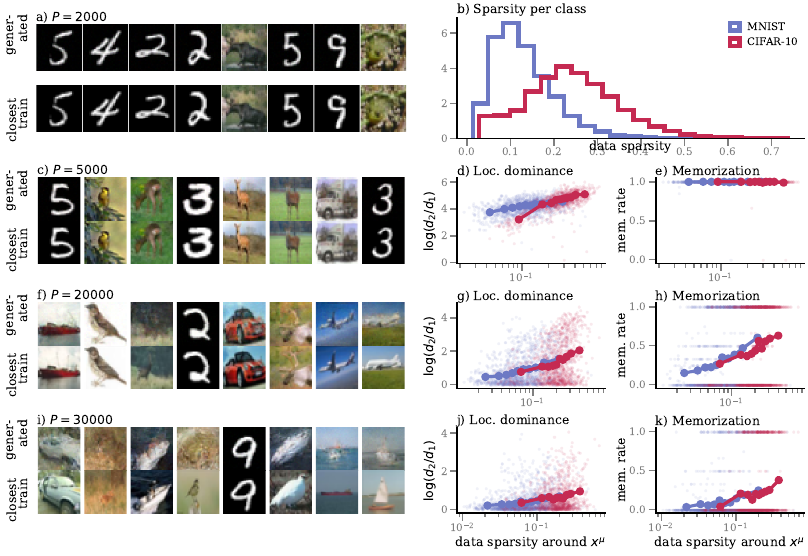}
    \caption{\textbf{Class-wise memorization on combined MNIST and CIFAR-10.}
    a) Generated samples together with their closest training example (measured by cosine similarity) for a diffusion model trained on $P=2000$ samples.
    b) Distribution of per-class sparsity around training points for four classes in CIFAR-10 measured from $20{,}000$ training samples.
    c) Generated samples together with their closest training example (measured by cosine similarity) for a diffusion model trained on $P=5000$ samples.
    d) Class-wise local dominance of a training point $x^\mu$ as a function of local data sparsity. Local dominance is defined via the ratio between the distance $d_1$ to the closest generated sample and the distance $d_2$ between that generated sample and the second-closest training point. MNIST points are shown in blue, CIFAR-10 points in red. 
    e) Memorization rate of each sample $x^\mu$, defined as the fraction of generated samples satisfying $d_1/d_2 < \tau = 1/3$. Different classes are shown as different colors.
    f)--k) Same measurements for models trained on larger datasets. As the number of training samples increases, both dominance and memorization decrease, and their dependence on sparsity weakens.
    }
    \label{fig:classwise_cifar_mnist}
\end{figure*}

In a separate experiment, we train diffusion models on a mixture of MNIST \citep{deng_mnist_2012} and CIFAR-10 images. MNIST, consisting of handwritten digits, is structurally simpler and occupies a much more concentrated region of image space, whereas CIFAR-10 exhibits substantially higher variability. Treating MNIST and CIFAR-10 as two distinct classes, we observe that MNIST samples consistently exhibit lower local sparsity than CIFAR-10 samples. As predicted by the theory, this translates into reduced memorization: in \cref{fig:classwise_cifar_mnist}, both local dominance and memorization rates increase with sparsity. For the comparison of CIFAR and MNIST, all distances are computed on $\ell_2$-normalized samples, so that sparsity reflects relative geometric structure (angular similarity) rather than differences in overall scale. We verified that using unnormalized samples yields qualitatively similar results. On average, MNIST points lie in denser regions and are therefore memorized less than CIFAR-10 points.

\section{Experimental Details}
\label{app: Details}
\paragraph{Model.}
We train denoising diffusion models using a standard discrete-time formulation with $T=1000$ timesteps. The score network is parameterized by a U-Net architecture \cite{ronneberger_u-net_2015} with convolutional filters. All models operate directly in pixel space.

\paragraph{Training procedure.}
Models are trained using the standard diffusion objective with mean squared error loss. Optimization is performed using Adam with learning rate $10^{-4}$ and batch size $100$. Training proceeds for a fixed number of $10^6$ gradient steps for all models. 

\paragraph{Evaluation}
For all experiments, the dataset is deterministically split into a training set of size $P$ and a held-out test set. The same split is reused for all evaluations to ensure comparability across checkpoints and metrics. 
At evaluation time, we generate $N_{\text{gen}}=4000$ samples from the trained diffusion model.
We compare generated samples to training and test data using either cosine similarity or normalized $\ell_2$ distance. In the cosine case, all inputs are flattened and $\ell_2$-normalized before computing similarities. 
Memorization is quantified at the level of individual generated samples. For each generated sample $\tilde{x}$, we compute its two nearest training neighbors and define distances $d_1$ and $d_2$. A sample is considered memorized if
$\frac{d_1}{d_2} < \tau $
for a range of thresholds $\tau \in \{1/6, 1/4, 1/3, 1/2, 2/3\}$. This yields both per-sample and per-training-point memorization statistics.
Local data density is estimated using nearest-neighbor statistics computed solely on the training set. For each training sample, we compute the distances to the $k$ nearest neighbors for $k \in \{2,5,10,20,50\}$ and average over these distances as defined in \cref{eq:sparsity_def}.
These statistics serve as proxies for local sparsity and are later correlated with memorization behavior. 

\subsection{CIFAR-10 Experiments}

We construct the CIFAR-10 dataset by combining the original training and test splits (60,000 images total), and then randomly partitioning them into three equal subsets (train/validation/test), each containing approximately 20,000 images.
All images are represented as $32 \times 32$ RGB tensors normalized to $[0,1]$. No additional preprocessing or augmentation is applied.

\subsection{MNIST + CIFAR-10 Experiments}

To study memorization under controlled differences in local data density, we construct a combined dataset from CIFAR-10 and MNIST, which exhibit markedly different intrinsic complexities. CIFAR-10 images are highly variable, while MNIST digits occupy a much more concentrated region of image space. This setup allows us to induce systematic differences in local sparsity across classes.

We merge the original training and test splits of both datasets. To ensure compatibility, MNIST images are resized to $32 \times 32$ and converted to three channels by duplicating the grayscale channel, resulting in RGB tensors consistent with CIFAR-10. CIFAR-10 images are used without modification. All images are represented as $32 \times 32$ RGB tensors normalized to $[0,1]$, and no additional preprocessing or augmentation is applied.

To isolate dataset-level effects, we ignore the original class labels and instead assign a binary label indicating the dataset of origin (CIFAR-10 or MNIST). The two datasets are then balanced by subsampling the larger dataset such that both contribute an equal number of samples. The resulting dataset therefore consists of an equal mixture of CIFAR-10 and MNIST images.

\subsection{CelebA Experiments}

This dataset consists of high-resolution face images together with $40$ binary attributes per image. All images are resized to $32 \times 32$ pixels, and converted to grayscale. The resulting images are represented as single-channel tensors normalized to $[0,1]$.
No additional data augmentation is applied.

\end{document}